\documentclass[10pt]{article}
\usepackage{amsfonts,amsmath,amssymb,cite}
\newcommand{\Real}{\mathop{\textrm{Re}}\nolimits}

\begin{document}
\title{Solution of the momentum-space Schr{\"o}dinger equation \\
for bound states of the $N$-dimensional Coulomb problem 
(revisited)}
\author{Rados{\l}aw Szmytkowski\footnote{Email: 
radek@mif.pg.gda.pl} \\*[3ex]
Atomic Physics Division,
Department of Atomic Physics and Luminescence, \\
Faculty of Applied Physics and Mathematics,
Gda{\'n}sk University of Technology, \\
Narutowicza 11/12, 80--233 Gda{\'n}sk, Poland}
\date{}
\maketitle
\begin{abstract}
The Schr{\"o}dinger--Coulomb Sturmian problem in $\mathbb{R}^{N}$,
$N\geqslant2$, is considered in the momentum representation. An
integral formula for the Gegenbauer polynomials, found recently by
Cohl [arXiv:1105.2735], is used to separate out angular variables and
reduce an integral Sturmian eigenvalue equation in $\mathbb{R}^{N}$
to a Fredholm one on $\mathbb{R}_{+}$. A kernel of the latter
equation contains the Legendre function of the second kind. A
symmetric Poisson-type series expansion of that function into
products of the Gegenbauer polynomials, established by Ossicini
[Boll.\ Un.\ Mat.\ Ital.\ 7 (1952) 315], is then used to determine
the Schr{\"o}dinger--Coulomb Sturmian eigenvalues and associated
momentum-space eigenfunctions. Finally, a relationship existing
between solutions to the Sturmian problem and solutions to a
(physically more interesting) energy eigenvalue problem is exploited
to find the Schr{\"o}dinger--Coulomb bound-state energy levels in
$\mathbb{R}^{N}$, together with explicit representations of the
associated normalized momentum-space Schr{\"o}dinger--Coulomb
Hamiltonian eigenfunctions.
\end{abstract}
%
%
\section{Introduction}
\label{I}
\setcounter{equation}{0}
A closed-form expression for the bound-state hydrogenic wave
functions in the momentum representation in $\mathbb{R}^{3}$ was
first obtained in 1929 by Podolsky and Pauling \cite{Podo29}, who
succeeded to carry out a Fourier transform of the position-space
Coulomb eigenfunctions, found three years earlier by Schr{\"o}dinger
\cite{Schr26}. Guided by the results of Ref.\ \cite{Podo29}, in 1932
Hylleraas \cite{Hyll32} derived a second-order partial differential
equation obeyed by the momentum-space Coulomb wave functions. In
1933, Elsasser \cite{Elsa33} attacked the problem in the spirit of
Ref.\ \cite{Podo29}, but with the Fourier transform done with the aid
of the complex contour-integration technique. A completely different
approach to the subject was presented in 1935 by Fock \cite{Fock35}.
He showed that the momentum-space Schr{\"o}dinger--Coulomb wave
equation in $\mathbb{R}^{3}$, known to be an \emph{integral\/} one,
might be transformed into a form which appeared to be identical with
an equation satisfied on the unit hypersphere $\mathbb{S}^{3}$ by the
hyperspherical harmonics. This fact led Fock to the conclusion that
for bound states the symmetry group of the quantum-mechanical Coulomb
problem in $\mathbb{R}^{3}$ is $O(4)$. That thread was pursued
further a year later by Bargmann \cite{Barg36}. Since then, the
problem was revisited and discussed in various aspects by a number of
authors. Space limitations allow us to mention here only several
representative papers \cite{Levy50,Shib65,Klei66,Pere66,Aldr67,%
Ivas72,deLa87,Hey93,Hols95,Aqui97,Ditt99,Aqui03,Mere06,Lin08}, as
well as relevant reviews and monographs \cite{Band66,Popo67,Chen72,%
vanH85,Aver89,Aver00,Aqui01,Aver03}, where references to other
pertinent works may be found in abundance.

In this paper, we present a method which allows one to determine
bound-state energy levels and associated Hamiltonian eigenfunctions
of the Schr{\"o}dinger--Coulomb problem in the $N$-dimensional
($N\geqslant2$) Euclidean momentum space, making no use of the
$O(N+1)$ dynamical symmetry specific to the problem nor of the
Erd{\'e}lyi's generalization \cite{Erde38} (cf.\ also Ref.\
\cite[Sec.\ 11.4]{Erde53}) of the Funk--Hecke theorem
\cite{Funk16,Heck17} regarding the integral equation obeyed by the
hyperspherical harmonics (which constituted the basis of the approach
presented by L{\'e}vy \cite{Levy50} in the case of $N=3$). We proceed
along the following steps. First, we replace the integral energy
eigenproblem by a related Sturmian integral eigenvalue equation, in
which energy is fixed and the role of the eigenparameter is played by
the Coulomb potential strength. Next, we reduce the aforementioned
Sturmian integral equation in $\mathbb{R}^{N}$ to a Fredholm one on
$\mathbb{R}_{+}$, exploiting a recent result from the theory of
Gegenbauer polynomials due to Cohl \cite{Cohl11}. A kernel in the
latter one-dimensional equation contains the Legendre function of the
second kind. We solve that integral equation using a Poisson-type
expansion of the aforementioned Legendre function in Gegenbauer
polynomials, derived long ago by Ossicini \cite{Ossi52}, but now
buried deeply in the mathematical literature and little known.
Finally, we use a relationship existing between solutions to the
Sturmian problem and to the energy eigenvalue problem, and determine
the Schr{\"o}dinger--Coulomb bound-state energy levels in
$\mathbb{R}^{N}$, together with explicit representations of the
associated normalized momentum-space Schr{\"o}dinger--Coulomb
Hamiltonian eigenfunctions.
%
%
\section{Integral equation for the momentum-space
Schr{\"o}dinger--Coulomb wave functions}
\label{II}
\setcounter{equation}{0}
In the position representation, the time-independent Schr{\"o}dinger
equation for a nonrelativistic spinless particle moving in the
$N$-dimensional ($N\geqslant2$) Euclidean space $\mathbb{R}^{N}$, in
the field of a force derivable from the scalar local potential
$V(\boldsymbol{r})$, is
\begin{equation}
-\frac{\hbar^{2}}{2m}\boldsymbol{\nabla}^{2}\Psi(\boldsymbol{r})
+V(\boldsymbol{r})\Psi(\boldsymbol{r})=E\Psi(\boldsymbol{r}).
\label{2.1}
\end{equation}
The counterpart equation in the momentum representation is
\begin{equation}
\left(\frac{p^{2}}{2m}-E\right)\Phi(\boldsymbol{p})
+\frac{1}{(2\pi\hbar)^{N/2}}
\int_{\mathbb{R}^{N}}\mathrm{d}^{N}\boldsymbol{p}'\:
U(\boldsymbol{p}-\boldsymbol{p}')
\Phi(\boldsymbol{p}')=0,
\label{2.2}
\end{equation}
where $\Phi(\boldsymbol{p})$ and $U(\boldsymbol{p})$ are the Fourier
transforms of $\Psi(\boldsymbol{r})$ and $V(\boldsymbol{r})$,
respectively. In the case of the attractive Coulomb potential
\begin{equation}
V(\boldsymbol{r})=-\frac{Ze^{2}}{(4\pi\epsilon_{0})r}
\qquad (Z>0),
\label{2.3}
\end{equation}
which concerns us here, $U(\boldsymbol{p})$ may be shown
(cf.\ Appendix) to be
\begin{equation}
U(\boldsymbol{p})=-\frac{Ze^{2}}{(4\pi\epsilon_{0})}
\frac{(2\hbar)^{N/2-1}\Gamma\left(\frac{N-1}{2}\right)}{\sqrt{\pi}}
\frac{1}{p^{N-1}},
\label{2.4}
\end{equation}
so that Eq.\ (\ref{2.2}) becomes
\begin{equation}
\left(\frac{p^{2}}{2m}-E\right)\Phi(\boldsymbol{p})
=\frac{Ze^{2}}{(4\pi\epsilon_{0})}
\frac{\Gamma\left(\frac{N-1}{2}\right)}{2\pi^{(N+1)/2}\hbar}
\int_{\mathbb{R}^{N}}\mathrm{d}^{N}\boldsymbol{p}'\:
\frac{\Phi(\boldsymbol{p}')}{|\boldsymbol{p}'-\boldsymbol{p}|^{N-1}}.
\label{2.5}
\end{equation}
We aim at finding solutions to the above equation which are
associated with, yet unknown, \emph{negative\/} (bound-state) energy
eigenvalues $E$.
%
%
\section{Solution of the integral equation for the momentum-space 
Schr{\"o}dinger--Coulomb Sturmian functions}
\label{III} 
\setcounter{equation}{0}
\subsection{Preliminaries}
\label{III.1}
Instead of attacking Eq.\ (\ref{2.5}) directly, it is easier to
consider first the associated Sturmian eigenproblem
\begin{equation}
\left(\frac{p^{2}}{2m}-E\right)\Sigma(E,\boldsymbol{p})
=\lambda\frac{Ze^{2}}{(4\pi\epsilon_{0})}
\frac{\Gamma\left(\frac{N-1}{2}\right)}{2\pi^{(N+1)/2}\hbar}
\int_{\mathbb{R}^{N}}\mathrm{d}^{N}\boldsymbol{p}'\:
\frac{\Sigma(E,\boldsymbol{p}')}
{|\boldsymbol{p}'-\boldsymbol{p}|^{N-1}},
\label{3.1}
\end{equation}
in which $\lambda\equiv\lambda(E)$ is an eigenvalue, whereas $E<0$ is
a \emph{fixed\/} parameter. It is evident that once the
$\lambda$-spectrum ($\{\lambda_{\alpha}\}$) for Eq.\ (\ref{3.1}) is
determined, the sought negative energy eigenvalues to the original
problem (\ref{2.5}) are roots to the algebraic equations
\begin{equation}
\lambda_{\alpha}(E)=1
\qquad (E<0).
\label{3.2}
\end{equation}
If $E_{\alpha\beta}<0$ is a particular solution to Eq.\ (\ref{3.2})
(the second subscript at $E$ serves to distinguish between different
roots to Eq.\ (\ref{3.2}), if there are more such roots than one),
the corresponding negative-energy momentum-space Coulomb wave
function $\Phi_{\alpha\beta\gamma}(\boldsymbol{p})$ is
\begin{equation}
\Phi_{\alpha\beta\gamma}(\boldsymbol{p})
=A\Sigma_{\alpha\gamma}(E_{\alpha\beta},\boldsymbol{p}),
\label{3.3}
\end{equation}
$A$ being an arbitrary non-zero constant (the second subscript at
$\Sigma$ appears if the eigenvalue $\lambda_{\alpha}$ is degenerate).

Before proceeding further, we shall rewrite Eq.\ (\ref{3.1}) in a
slightly more compact form. To this end, we define
\begin{equation}
q=\sqrt{-2mE},
\qquad
q_{\mathrm{B}}=\frac{\hbar}{a_{\mathrm{B}}}
\label{3.4}
\end{equation}
(here $a_{\mathrm{B}}=(4\pi\epsilon_{0})\hbar^{2}/me^{2}$ is the Bohr
radius). Use of Eq.\ (\ref{3.4}) transforms Eq.\ (\ref{3.1}) into
\begin{equation}
(p^{2}+q^{2})\Sigma(E,\boldsymbol{p})=\lambda Zq_{\mathrm{B}}
\frac{\Gamma\left(\frac{N-1}{2}\right)}{\pi^{(N+1)/2}}
\int_{\mathbb{R}^{N}}\mathrm{d}^{N}\boldsymbol{p}'\:
\frac{\Sigma(E,\boldsymbol{p}')}
{|\boldsymbol{p}'-\boldsymbol{p}|^{N-1}}.
\label{3.5}
\end{equation}
\subsection{Reduction of Eq.\ (\ref{3.5}) to the radial form}
\label{III.2}
We try whether Eq.\ (\ref{3.5}) has solutions in the form
\begin{equation}
\Sigma_{l\eta}(E,\boldsymbol{p})=F_{l}(E,p)
Y_{l\eta}^{(N-1)}(\boldsymbol{n}_{p}).
\label{3.6}
\end{equation}
Here $Y_{l\eta}^{(N-1)}(\boldsymbol{n}_{p})$, with $l\in\mathbb{N}$
and $\eta\in\{1,2,\ldots,d_{l}^{(N-1)}\}$, where
\begin{equation}
d_{l}^{(N-1)}=(2l+N-2)
\frac{(l+N-3)!}{l!(N-2)!},
\label{3.7}
\end{equation}
are the hyperspherical harmonics \cite{Aver89,Aver00}, that are
orthonormal on the unit hypersphere $\mathbb{S}^{N-1}$ in the sense
of
\begin{equation}
\oint_{\mathbb{S}^{N-1}}\mathrm{d}^{N-1}\boldsymbol{n}_{p}\:
Y_{l\eta}^{(N-1)*}(\boldsymbol{n}_{p})
Y_{l'\eta'}^{(N-1)}(\boldsymbol{n}_{p})
=\delta_{ll'}\delta_{\eta\eta'}.
\label{3.8}
\end{equation}
We insert Eq.\ (\ref{3.6}) into Eq.\ (\ref{3.5}), multiply the
resulting equation by $Y_{l\eta}^{(N-1)*}(\boldsymbol{n}_{p})$, and
then sum over $\eta$. Using the addition theorem \cite[Eqs.\ (3.80),
(3.86) and (3.40)]{Aver00}
\begin{equation}
\sum_{\eta=1}^{d_{l}^{(N-1)}}Y_{l\eta}^{(N-1)*}(\boldsymbol{n}_{p})
Y_{l\eta}^{(N-1)}(\boldsymbol{n}_{p}^{\prime})
=\frac{d_{l}^{(N-1)}}{S_{N-1}}\frac{C_{l}^{(N/2-1)}(\boldsymbol{n}_{p}
\cdot\boldsymbol{n}_{p}^{\prime})}{C_{l}^{(N/2-1)}(1)},
\label{3.9}
\end{equation}
where $C_{n}^{(\alpha)}(\xi)$ is the Gegenbauer polynomial and
$S_{N-1}$ is the surface area of $\mathbb{S}^{N-1}$, we obtain
\begin{eqnarray}
(p^{2}+q^{2})F_{l}(E,p)
&=& \lambda Zq_{\mathrm{B}}\frac{\Gamma\left(\frac{N-1}{2}\right)}
{2^{(N-1)/2}\pi^{(N+1)/2}C_{l}^{(N/2-1)}(1)p^{(N-1)/2}}
\int_{0}^{\infty}\mathrm{d}p'\:p^{\prime\,(N-1)/2}F_{l}(E,p')
\nonumber \\
&& \times\oint_{\mathbb{S}^{N-1}}
\mathrm{d}^{N-1}\boldsymbol{n}_{p}^{\prime}\:
\frac{C_{l}^{(N/2-1)}(\boldsymbol{n}_{p}
\cdot\boldsymbol{n}_{p}^{\prime})}
{\displaystyle\left(\frac{p^{2}+p^{\prime\,2}}{2pp'}
-\boldsymbol{n}_{p}\cdot\boldsymbol{n}_{p}^{\prime}\right)^{(N-1)/2}}.
\label{3.10}
\end{eqnarray}
It is evident that the angular integral appearing in Eq.\
(\ref{3.10}) may be expressed as
\begin{equation}
\oint_{\mathbb{S}^{N-1}}\mathrm{d}^{N-1}\boldsymbol{n}_{p}^{\prime}\:
\frac{C_{l}^{(N/2-1)}(\boldsymbol{n}_{p}
\cdot\boldsymbol{n}_{p}^{\prime})}
{\displaystyle\left(\frac{p^{2}+p^{\prime\,2}}{2pp'}
-\boldsymbol{n}_{p}\cdot\boldsymbol{n}_{p}^{\prime}\right)^{(N-1)/2}}
=S_{N-2}\int_{-1}^{1}\mathrm{d}\xi\:(1-\xi^2)^{(N-3)/2}
\frac{C_{l}^{(N/2-1)}(\xi)}{\displaystyle
\left(\frac{p^{2}+p^{\prime\,2}}{2pp'}-\xi\right)^{(N-1)/2}}.
\label{3.11}
\end{equation}
However, it has been recently shown by Cohl \cite{Cohl11} that
\begin{eqnarray}
&& \int_{-1}^{1}\mathrm{d}\xi\:\frac{(1-\xi^{2})^{\alpha-1/2}
C_{n}^{(\alpha)}(\xi)}{(z-\xi)^{\kappa+1/2}}
=\mathrm{e}^{\mathrm{i}\pi(\alpha-\kappa)}
2^{\alpha+1/2}C_{n}^{(\alpha)}(1)
\frac{\Gamma(\alpha+\frac{1}{2})}{\Gamma(\kappa+\frac{1}{2})}
(z^{2}-1)^{(\alpha-\kappa)/2}
\mathfrak{Q}_{n+\alpha-1/2}^{\kappa-\alpha}(z)
\nonumber \\
&& \hspace*{15em} (\textrm{$\Real\alpha>{\textstyle-\frac{1}{2}}$,
$\kappa\in\mathbb{C}$, $n\in\mathbb{N}$, 
$z\in\mathbb{C}\setminus(-\infty,1]$}),
\label{3.12}
\end{eqnarray}
where $\mathfrak{Q}_{\nu}^{\mu}(z)$ is the associated Legendre
function of the second kind. It follows from Eq.\ (\ref{3.12}) that
\begin{eqnarray}
\int_{-1}^{1}\mathrm{d}\xi\:\frac{(1-\xi^{2})^{\alpha-1/2}
C_{n}^{(\alpha)}(\xi)}{(z-\xi)^{\alpha+1/2}}
&=& 2^{\alpha+1/2}C_{n}^{(\alpha)}(1)\mathfrak{Q}_{n+\alpha-1/2}(z)
\nonumber \\
&& \hspace*{3em} (\textrm{$\Real\alpha>{\textstyle-\frac{1}{2}}$,
$n\in\mathbb{N}$, $z\in\mathbb{C}\setminus(-\infty,1]$})
\label{3.13}
\end{eqnarray}
($\mathfrak{Q}_{\nu}(z)\equiv\mathfrak{Q}_{\nu}^{0}(z)$ is the
Legendre function of the second kind), and consequently
\begin{equation}
\oint_{\mathbb{S}^{N-1}}\mathrm{d}^{N-1}\boldsymbol{n}_{p}^{\prime}\:
\frac{C_{l}^{(N/2-1)}(\boldsymbol{n}_{p}
\cdot\boldsymbol{n}_{p}^{\prime})}
{\displaystyle\left(\frac{p^{2}+p^{\prime\,2}}{2pp'}
-\boldsymbol{n}_{p}\cdot\boldsymbol{n}_{p}^{\prime}\right)^{(N-1)/2}}
=2^{(N-1)/2}S_{N-2}C_{l}^{(N/2-1)}(1)\mathfrak{Q}_{l+(N-3)/2}
\left(\frac{p^{2}+p^{\prime\,2}}{2pp'}\right).
\label{3.14}
\end{equation}
Plugging Eq.\ (\ref{3.14}) into Eq.\ (\ref{3.10}) and making use of
the relation
\begin{equation}
S_{N-2}=\frac{(N-1)\pi^{(N-1)/2}}{\Gamma(\frac{N+1}{2})},
\label{3.15}
\end{equation}
we arrive at the following integral equation for the radial function
$F_{l}(E,p)$:
\begin{equation}
(p^{2}+q^{2})F_{l}(E,p)=\lambda
\frac{2Zq_{\mathrm{B}}}{\pi p^{(N-1)/2}}
\int_{0}^{\infty}\mathrm{d}p'\:p^{\prime\,(N-1)/2}
\mathfrak{Q}_{l+(N-3)/2}
\left(\frac{p^{2}+p^{\prime\,2}}{2pp'}\right)F_{l}(E,p')
\label{3.16}
\end{equation}
(for $N=3$ and $\lambda=1$, Eq.\ (\ref{3.16}) coincides, as it
should, with Eq.\ (8.7) in Ref.\ \cite{Beth57}). After some
straightforward rearrangements, this may be further cast into the
form
\begin{eqnarray}
&& \hspace*{-2em}
[\sqrt{p^{2}+q^{2}}\,p^{(N-1)/2}F_{l}(E,p)]
\nonumber \\
&& =\,\lambda\frac{2Zq_{\mathrm{B}}}{\pi}
\int_{0}^{\infty}\mathrm{d}p'\:
\frac{\displaystyle\mathfrak{Q}_{l+(N-3)/2}
\left(\frac{p^{2}+p^{\prime\,2}}{2pp'}\right)}
{\sqrt{p^{2}+q^{2}}\sqrt{p^{\prime\,2}+q^{2}}}
[\sqrt{p^{\prime\,2}+q^{2}}\,p^{\prime\,(N-1)/2}F_{l}(E,p')],
\label{3.17}
\end{eqnarray}
the kernel in the above equation being real and manifestly symmetric.
\subsection{Intermezzo on the spectral theory for Fredholm integral
equations}
\label{III.3}
Before we attack Eq.\ (\ref{3.17}), we have to recall some very basic
facts from the spectral theory for integral equations. If the kernel
$M(p,p')$ in the homogeneous Fredholm equation of the second kind
\begin{equation}
f(p)=\lambda\int_{0}^{\infty}\mathrm{d}p'\:M(p,p')f(p')
\label{3.18}
\end{equation}
is real and symmetric, and if its $\lambda$-spectrum (known to be
real, in view of the aforementioned assumptions about the kernel) is
purely discrete, the following spectral expansion of $M(p,p')$ holds:
\begin{equation}
M(p,p')=\sum_{k}\frac{f_{k}(p)f_{k}(p')}{\lambda_{k}},
\label{3.19}
\end{equation}
provided the eigenfunctions $\{f_{k}(p)\}$ are orthonormal in the
sense of
\begin{equation}
\int_{0}^{\infty}\mathrm{d}p\:f_{k}(p)f_{k'}(p)=\delta_{kk'}.
\label{3.20}
\end{equation}
This implies that if we cope with some particular equation of the
form (\ref{3.18}), and if, in whatever manner and from whichever
premises, we infer that the kernel $M(p,p')$ possesses the series
representation
\begin{equation}
M(p,p')=\sum_{k}\varepsilon_{k}g_{k}(p)g_{k}(p')
\qquad (\varepsilon_{k}=\pm1),
\label{3.21}
\end{equation}
the functions $\{g_{k}(p)\}$ being orthogonal,
\begin{equation}
\int_{0}^{\infty}\mathrm{d}p\:g_{k}(p)g_{k'}(p)
=||g_{k}||^{2}\delta_{kk'},
\label{3.22}
\end{equation}
we may immediately deduce that the eigenvalues for Eq.\ (\ref{3.18})
are
\begin{equation}
\lambda_{k}=\varepsilon_{k}||g_{k}||^{-2},
\label{3.23}
\end{equation}
whereas the associated normalized eigenfunctions are
\begin{equation}
f_{k}(p)=\frac{g_{k}(p)}{||g_{k}||}
=\sqrt{\varepsilon_{k}\lambda_{k}}\,g_{k}(p)
\label{3.24}
\end{equation}
(observe that $\varepsilon_{k}^{-1}=\varepsilon_{k}$).
\subsection{Solution of the radial integral equation (\ref{3.17})}
\label{III.4}
We return to Eq.\ (\ref{3.17}). It is evident that it falls into the
category (\ref{3.18}), with
\begin{equation}
f_{l}(p)=\sqrt{p^{2}+q^{2}}\,p^{(N-1)/2}F_{l}(E,p)
\label{3.25}
\end{equation}
(here the subscript $l$ at $f$ corresponds to the one at $F$) and
\begin{equation}
M_{l}(p,p')=\frac{2Zq_{\mathrm{B}}}{\pi}
\frac{\displaystyle\mathfrak{Q}_{l+(N-3)/2}
\left(\frac{p^{2}+p^{\prime\,2}}{2pp'}\right)}
{\sqrt{p^{2}+q^{2}}\sqrt{p^{\prime\,2}+q^{2}}}.
\label{3.26}
\end{equation}
An expansion of the kernel (\ref{3.26}) in the symmetric form
(\ref{3.21}) may be derived from the following Poisson-type identity:
\begin{eqnarray}
\mathfrak{Q}_{\nu}\left(\frac{1-2h\xi\xi'+h^{2}}
{2h\sqrt{(1-\xi^{2})(1-\xi^{\prime\,2})}}\right)
&=& 2^{2\nu+1}[\Gamma(\nu+1)]^{2}
\left(1-\xi^{2}\right)^{(\nu+1)/2}
\left(1-\xi^{\prime\,2}\right)^{(\nu+1)/2}
\nonumber \\
&& \times\sum_{n_{\mathrm{r}}=0}^{\infty}
\frac{n_{\mathrm{r}}!}{\Gamma(n_{\mathrm{r}}+2\nu+2)}\,
h^{n_{\mathrm{r}}+\nu+1}C_{n_{\mathrm{r}}}^{(\nu+1)}(\xi)
C_{n_{\mathrm{r}}}^{(\nu+1)}(\xi')
\nonumber \\
&& \hspace*{10em} (\textrm{$|h|\leqslant1$, 
$-1\leqslant\xi,\xi'\leqslant1$}),
\label{3.27}
\end{eqnarray}
discovered several decades ago by Ossicini \cite[Eq.\ (11)]{Ossi52}
(there is a misprint in the formula provided originally by Ossicini
--- on its left-hand side $[\Gamma(\lambda)]$ should be replaced by
$[\Gamma(\lambda)]^{2}$; moreover, one should be aware that in Ref.\
\cite[Eq.\ (47.6.11)]{Hans75} the formula in question was reprinted
with several errors!). For $h=1$, Eq.\ (\ref{3.27}) becomes
\begin{eqnarray}
\mathfrak{Q}_{\nu}\left(\frac{1-\xi\xi'}
{\sqrt{(1-\xi^{2})(1-\xi^{\prime\,2})}}\right)
&=& 2^{2\nu+1}[\Gamma(\nu+1)]^{2}
\left(1-\xi^{2}\right)^{(\nu+1)/2}
\left(1-\xi^{\prime\,2}\right)^{(\nu+1)/2}
\nonumber \\
&& \times\sum_{n_{\mathrm{r}}=0}^{\infty}
\frac{n_{\mathrm{r}}!}{\Gamma(n_{\mathrm{r}}+2\nu+2)}\,
C_{n_{\mathrm{r}}}^{(\nu+1)}(\xi)C_{n_{\mathrm{r}}}^{(\nu+1)}(\xi')
\nonumber \\
&& \hspace*{12em} (-1\leqslant\xi,\xi'\leqslant1),
\label{3.28}
\end{eqnarray}
from which, with the substitutions
\begin{equation}
\xi=\frac{q^{2}-p^{2}}{q^{2}+p^{2}},
\qquad
\xi'=\frac{q^{2}-p^{\prime\,2}}{q^{2}+p^{\prime\,2}},
\label{3.29}
\end{equation}
we infer that
\begin{eqnarray}
\mathfrak{Q}_{\nu}\left(\frac{p^{2}+p^{\prime\,2}}
{2pp^{\prime}}\right) &=& 2^{2\nu+1}[\Gamma(\nu+1)]^{2}
\left(\frac{2qp}{q^{2}+p^{2}}\right)^{\nu+1}
\left(\frac{2qp'}{q^{2}+p^{\prime\,2}}\right)^{\nu+1}
\nonumber \\
&& \times\sum_{n_{\mathrm{r}}=0}^{\infty}
\frac{n_{\mathrm{r}}!}{\Gamma(n_{\mathrm{r}}+2\nu+2)}\,
C_{n_{\mathrm{r}}}^{(\nu+1)}
\left(\frac{q^{2}-p^{2}}{q^{2}+p^{2}}\right)
C_{n_{\mathrm{r}}}^{(\nu+1)}
\left(\frac{q^{2}-p^{\prime\,2}}{q^{2}+p^{\prime\,2}}\right)
\label{3.30}
\end{eqnarray}
(the particular reason for which we have denoted the summation index
as $n_{\mathrm{r}}$ will become clear shortly). Combining Eqs.\
(\ref{3.26}) and (\ref{3.30}), we see that the kernel in Eq.\
(\ref{3.17}) may be written as
\begin{equation}
M_{l}(p,p')=\sum_{n_{\mathrm{r}}=0}^{\infty}
g_{n_{\mathrm{r}}l}(p)g_{n_{\mathrm{r}}l}(p')
\qquad (\Rightarrow\forall\,n_{\mathrm{r}}\in\mathbb{N}:\:
\varepsilon_{n_{\mathrm{r}}l}=1),
\label{3.31}
\end{equation}
with
\begin{equation}
g_{n_{\mathrm{r}}l}(p)=\Gamma\left(l+\frac{N-1}{2}\right)
\sqrt{\frac{Zq_{\mathrm{B}}n_{\mathrm{r}}!}
{\pi(n_{\mathrm{r}}+2l+N-2)!}}\,
\frac{(4qp)^{l+(N-1)/2}}{(q^{2}+p^{2})^{l+N/2}}\,
C_{n_{\mathrm{r}}}^{(l+(N-1)/2)}
\left(\frac{q^{2}-p^{2}}{q^{2}+p^{2}}\right).
\label{3.32}
\end{equation}
In the next step, we evaluate the integral
$\int_{0}^{\infty}\mathrm{d}p\:g_{n_{\mathrm{r}}l}(p)
g_{n_{\mathrm{r}}^{\prime}l}(p)$, in order to verify whether the
functions (\ref{3.32}) form an orthogonal set [as they should, cf.\
Eq.\ (\ref{3.22})]. Exploiting the known \cite[Eq.\ (7.313)]{Grad07}
integral formula
\begin{equation}
\int_{-1}^{1}\mathrm{d}\xi\:(1-\xi^{2})^{\alpha-1/2}
C_{n_{\mathrm{r}}}^{(\alpha)}(\xi)
C_{n_{\mathrm{r}}^{\prime}}^{(\alpha)}(\xi)
=\frac{\pi\Gamma(n_{\mathrm{r}}+2\alpha)}
{2^{2\alpha-1}n_{\mathrm{r}}!
(n_{\mathrm{r}}+\alpha)[\Gamma(\alpha)]^{2}}\,
\delta_{n_{\mathrm{r}}n_{\mathrm{r}}^{\prime}}
\qquad (\textrm{$\Real\alpha>{\textstyle-\frac{1}{2}}$, 
$\alpha\neq0$}),
\label{3.33}
\end{equation}
we find that
\begin{equation}
\int_{0}^{\infty}\mathrm{d}p\:
g_{n_{\mathrm{r}}l}(p)g_{n_{\mathrm{r}}^{\prime}l}(p)
=\frac{Z}{n_{\mathrm{r}}+l+\frac{N-1}{2}}\frac{q_{\mathrm{B}}}{q}\,
\delta_{n_{\mathrm{r}}n_{\mathrm{r}}^{\prime}},
\label{3.34}
\end{equation}
i.e., the functions (\ref{3.32}) are indeed orthogonal, the norm of
$g_{n_{\mathrm{r}}l}(p)$ being
\begin{equation}
||g_{n_{\mathrm{r}}l}||
=\sqrt{\frac{Z}{n_{\mathrm{r}}+l+\frac{N-1}{2}}
\frac{q_{\mathrm{B}}}{q}}.
\label{3.35}
\end{equation}
Hence, applying Eqs.\ (\ref{3.23}) and (\ref{3.24}) we infer that the
spectrum of eigenvalues in Eq.\ (\ref{3.17}) is given by
\begin{equation}
\lambda_{n_{\mathrm{r}}l}=\frac{n_{\mathrm{r}}+l+\frac{N-1}{2}}{Z}
\frac{q}{q_{\mathrm{B}}}
\qquad (n_{\mathrm{r}}\in\mathbb{N}),
\label{3.36}
\end{equation}
whereas the associated orthonormal eigenfunctions are
\begin{eqnarray}
f_{n_{\mathrm{r}}l}(E,p) &=& \Gamma\left(l+\frac{N-1}{2}\right)
\sqrt{\frac{qn_{\mathrm{r}}!(n_{\mathrm{r}}+l+\frac{N-1}{2})}
{\pi(n_{\mathrm{r}}+2l+N-2)!}}
\nonumber \\
&& \times\,\frac{(4qp)^{l+(N-1)/2}}{(q^{2}+p^{2})^{l+N/2}}\,
C_{n_{\mathrm{r}}}^{(l+(N-1)/2)}
\left(\frac{q^{2}-p^{2}}{q^{2}+p^{2}}\right).
\label{3.37}
\end{eqnarray}
Consequently (cf.\ Eq.\ (\ref{3.25})), the sought radial
momentum-space Coulomb Sturmian functions are found to be
\begin{eqnarray}
F_{n_{\mathrm{r}}l}(E,p) 
&=& 2^{N-1}\Gamma\left(l+\frac{N-1}{2}\right)
\sqrt{\frac{n_{\mathrm{r}}!(n_{\mathrm{r}}+l+\frac{N-1}{2})}
{\pi(n_{\mathrm{r}}+2l+N-2)!}}
\nonumber \\
&& \times\,q^{N/2}\frac{(4qp)^{l}}{(q^{2}+p^{2})^{l+(N+1)/2}}\,
C_{n_{\mathrm{r}}}^{(l+(N-1)/2)}
\left(\frac{q^{2}-p^{2}}{q^{2}+p^{2}}\right)
\label{3.38}
\end{eqnarray}
(it is seen that $n_{\mathrm{r}}$ is simply the radial quantum
number). They obey the weighted orthonormality relation
\begin{equation}
\int_{0}^{\infty}\mathrm{d}p\:p^{N-1}(p^{2}+q^{2})
F_{n_{\mathrm{r}}l}(E,p)F_{n_{\mathrm{r}}^{\prime}l}(E,p)
=\delta_{n_{\mathrm{r}}n_{\mathrm{r}}^{\prime}}
\label{3.39}
\end{equation}
and are standardized to be positive in the immediate vicinity of
$p=0$.
\subsection{Some properties of solutions to Eq.\ (\ref{3.5})}
\label{III.5}
Once the radial integral equation (\ref{3.16}) is solved, we may
return to the $N$-dimensional problem (\ref{3.5}). 

It follows from Eq.\ (\ref{3.36}) that the degree of degeneracy of
the eigenvalues in Eq.\ (\ref{3.5}) is higher than $d_{l}^{(N-1)}$,
as $\lambda_{n_\mathrm{r}l}$ depends on the quantum numbers
$n_{\mathrm{r}}$ and $l$ through their sum. To distinguish between
different eigenvalues, we introduce the principal quantum number
\begin{equation}
n=n_{\mathrm{r}}+l+1,
\label{3.40}
\end{equation}
in terms of which we have
\begin{equation}
\lambda_{n_{\mathrm{r}}l}\equiv\lambda_{n}
=\frac{n+\frac{N-3}{2}}{Z}\frac{q}{q_{\mathrm{B}}}.
\label{3.41}
\end{equation}
The degree of degeneracy of the particular eigenvalue $\lambda_{n}$
obviously is
\begin{equation}
D_{n}^{(N)}=\sum_{l=0}^{n-1}d_{l}^{(N-1)}.
\label{3.42}
\end{equation}
In view of Eq.\ (\ref{3.7}), this may be rewritten as
\begin{equation}
D_{n}^{(N)}=2\sum_{l=0}^{n-1}{l+N-2 \choose N-2}
-\sum_{l=0}^{n-1}{l+N-3 \choose N-3}.
\label{3.43}
\end{equation}
The sums in Eq.\ (\ref{3.43}) are elementary; we have
\cite[Eq.\ (0.15.1)]{Grad07}
\begin{equation}
\sum_{k=0}^{n}{k+m \choose m}={n+m+1 \choose m+1},
\label{3.44}
\end{equation}
and consequently
\begin{equation}
D_{n}^{(N)}=(2n+N-3)\frac{(n+N-3)!}{(n-1)!(N-1)!}.
\label{3.45}
\end{equation}
Comparison of Eqs.\ (\ref{3.45}) and (\ref{3.7}) shows that it holds
that
\begin{equation}
D_{n}^{(N)}=d_{n-1}^{(N)}.
\label{3.46}
\end{equation}
(The relation in Eq.\ (\ref{3.46}) is, of course, a symptom of the
$O(N+1)$ dynamical symmetry of the $N$-dimensional negative-energy
Schr{\"o}dinger--Coulomb problem.)

Degenerate Sturmian eigenfunctions associated with the characteristic
numbers (\ref{3.41}), labeled with the principal rather than the
radial quantum number, are
\begin{equation}
\Sigma_{nl\eta}(E,\boldsymbol{p})
=F_{n-l-1,l}(E,p)Y_{l\eta}^{(N-1)}(\boldsymbol{n}_{p}).
\label{3.47}
\end{equation}
It is implied by Eqs.\ (\ref{3.8}) and (\ref{3.39}) that these
functions obey the weighted orthonormality relation
\begin{equation}
\int_{\mathbb{R}^{N}}\mathrm{d}^{N}\boldsymbol{p}\:
(p^{2}+q^{2})\Sigma_{nl\eta}^{*}(E,\boldsymbol{p})
\Sigma_{n'l'\eta'}(E,\boldsymbol{p})
=\delta_{nn'}\delta_{ll'}\delta_{\eta\eta'}.
\label{3.48}
\end{equation}

It remains to consider the completeness problem for the Sturmian set
$\{\Sigma_{nl\eta}(E,\boldsymbol{p})\}$. It is known that the
functions $\{(1-\xi^{2})^{\alpha/2-1/4}
C_{n_{\mathrm{r}}}^{(\alpha)}(\xi)\}$, with
$\Real\alpha>-\frac{1}{2}$, form a complete set in the functional
space $L^{2}((-1,1))$, the corresponding closure relation being
\begin{eqnarray}
\frac{2^{2\alpha-1}[\Gamma(\alpha)]^{2}}{\pi}
\sum_{n_{\mathrm{r}}=0}^{\infty}
\frac{n_{\mathrm{r}}!(n_{\mathrm{r}}+\alpha)}
{\Gamma(n_{\mathrm{r}}+2\alpha)}\,
C_{n_{\mathrm{r}}}^{(\alpha)}(\xi)
C_{n_{\mathrm{r}}}^{(\alpha)}(\xi') &=& \frac{\delta(\xi-\xi')}
{[(1-\xi^{2})(1-\xi^{\prime\,2})]^{\alpha/2-1/4}}
\nonumber \\
&& (\textrm{$-1<\xi,\xi'<1$,
$\Real\alpha>{\textstyle-\frac{1}{2}}$}),
\label{3.49}
\end{eqnarray}
with $\delta(\xi-\xi')$ denoting the Dirac delta distribution. We set
in the above equation $\alpha=l+(N-1)/2$, plug Eq.\ (\ref{3.29}),
combine the result with Eq.\ (\ref{3.38}), and then exploit the
identity
\begin{equation}
\delta(\xi-\xi')=\frac{(p^{2}+q^{2})(p^{\prime\,2}+q^{2})}
{4\sqrt{pp'}\,q^{2}}\,\delta(p-p'),
\label{3.50}
\end{equation}
which follows from Eq.\ (\ref{3.29}) and the well-known properties of
the Dirac delta. This chain of operations transforms Eq.\
(\ref{3.49}) into the closure relation for the radial Sturmians:
\begin{equation}
\sum_{n_{\mathrm{r}}=0}^{\infty}F_{n_{\mathrm{r}}l}(E,p)
F_{n_{\mathrm{r}}l}(E,p')=\frac{\delta(p-p')}
{(pp')^{(N-1)/2}\sqrt{(p^{2}+q^{2})(p^{\prime\,2}+q^{2})}}.
\label{3.51}
\end{equation}
Next, we have the closure relation
\begin{equation}
\sum_{l=0}^{\infty}\sum_{\eta=1}^{d_{l}^{(N-1)}}
Y_{l\eta}^{(N-1)}(\boldsymbol{n}_{p})
Y_{l\eta}^{(N-1)*}(\boldsymbol{n}_{p}^{\prime})
=\delta^{(N-1)}(\boldsymbol{n}_{p}-\boldsymbol{n}_{p}^{\prime}),
\label{3.52}
\end{equation}
which reflects the fact that the hyperspherical harmonics form a
complete set in the functional space $L^{2}(\mathbb{S}^{N-1})$. We
insert Eqs.\ (\ref{3.51}) and (\ref{3.52}) into the right-hand side
of the identity
\begin{equation}
\sum_{n=1}^{\infty}\sum_{l=0}^{n-1}\sum_{\eta=1}^{d_{l}^{(N-1)}}
\Sigma_{nl\eta}(E,\boldsymbol{p})
\Sigma_{nl\eta}^{*}(E,\boldsymbol{p}')
=\sum_{l=0}^{\infty}\sum_{\eta=1}^{d_{l}^{(N-1)}}
Y_{l\eta}^{(N-1)}(\boldsymbol{n}_{p})
Y_{l\eta}^{(N-1)*}(\boldsymbol{n}_{p}^{\prime})
\sum_{n_{\mathrm{r}}=0}^{\infty}F_{n_{\mathrm{r}}l}(E,p)
F_{n_{\mathrm{r}}l}(E,p')
\label{3.53}
\end{equation}
and make use of the obvious fact that
\begin{equation}
\delta^{(N)}(\boldsymbol{p}-\boldsymbol{p}')
=\frac{\delta(p-p')\delta^{(N-1)}
(\boldsymbol{n}_{p}-\boldsymbol{n}_{p}^{\prime})}{(pp')^{(N-1)/2}}.
\label{3.54}
\end{equation}
This leads us to the symmetric closure relation
\begin{equation}
\sum_{n=1}^{\infty}\sum_{l=0}^{n-1}\sum_{\eta=1}^{d_{l}^{(N-1)}}
\Sigma_{nl\eta}(E,\boldsymbol{p})
\Sigma_{nl\eta}^{*}(E,\boldsymbol{p}')
=\frac{\delta^{(N)}(\boldsymbol{p}-\boldsymbol{p}')}
{\sqrt{(p^{2}+q^{2})(p^{\prime\,2}+q^{2})}}.
\label{3.55}
\end{equation}
We have thus proved that the Sturmian functions
$\{\Sigma_{nl\eta}(E,\boldsymbol{p})\}$ form a complete orthonormal
set in the functional space $L_{p^{2}+q^{2}}^{2}(\mathbb{R}^{N})$;
equivalently, the functions $\{\sqrt{p^{2}+q^{2}}\,
\Sigma_{nl\eta}(E,\boldsymbol{p})\}$ form a complete orthonormal set
in the functional space $L^{2}(\mathbb{R}^{N})$.
%
%
\section{Determination of bound-state eigensolutions to the
momentum-space Schr{\"o}dinger--Coulomb wave equation (\ref{2.5})}
\label{IV}
\setcounter{equation}{0}
We are now ready to determine bound-state eigensolutions to Eq.\
(\ref{2.5}), following the procedure outlined at the beginning of
Sec.\ \ref{III.1}. At first, we substitute $\lambda_{n}$, as given in
Eq.\ (\ref{3.41}), for $\lambda_{\alpha}(E)$ in Eq.\ (\ref{3.2}),
then use Eq.\ (\ref{3.4}), and solve the resulting equation for $E$.
For each particular $n$, there is just one solution to that equation,
which we shall denote as $E_{n}$; it is
\begin{equation}
E_{n}=-\frac{Z^{2}e^{2}}
{2\left(n+\frac{N-3}{2}\right)^{2}(4\pi\epsilon_{0})a_{\mathrm{B}}}.
\label{4.1}
\end{equation}
Corresponding eigenfunctions, in accordance with Eqs.\ (\ref{3.3})
and (\ref{3.6}), are
\begin{equation}
\Phi_{nl\eta}(\boldsymbol{p})=\mathcal{F}_{nl}(p)
Y_{l\eta}^{(N-1)}(\boldsymbol{n}_{p}),
\label{4.2}
\end{equation}
with
\begin{equation}
\mathcal{F}_{nl}(p)=AF_{n-l-1,l}(E_{n},p).
\label{4.3}
\end{equation}
Equation (\ref{2.5}) shows that $E_{n}^{-1}$ is an eigenvalue of the
Hermitian (in fact --- real and symmetric) kernel
\begin{displaymath}
\frac{pp'}{2m}\delta^{(N)}(\boldsymbol{p}-\boldsymbol{p}')
-\frac{Ze^{2}}{(4\pi\epsilon_{0})}
\frac{\Gamma\left(\frac{N-1}{2}\right)}{2\pi^{(N+1)/2}\hbar}
|\boldsymbol{p}-\boldsymbol{p}'|^{-(N-1)}.
\end{displaymath}
From this and from Eqs.\ (\ref{4.2}) and (\ref{3.8}), one routinely
deduces that the functions $\{\Phi_{nl\eta}(\boldsymbol{p})\}$ are
orthogonal in the sense of
\begin{equation}
\int_{\mathbb{R}^{N}}\mathrm{d}^{N}\boldsymbol{p}\:
\Phi_{nl\eta}^{*}(\boldsymbol{p})\Phi_{n'l'\eta'}(\boldsymbol{p})
\sim\delta_{nn'}\delta_{ll'}\delta_{\eta\eta'}.
\label{4.4}
\end{equation}
If, additionally, we require the functions under consideration are
normalized to unity under the same scalar product that is used in the
above equation, and if we choose the constant $A$ in Eq.\ (\ref{4.3})
to be real, we obtain the constraint
\begin{equation}
A^{2}\int_{0}^{\infty}\mathrm{d}p\:p^{N-1}
[F_{n-l-1,l}(E_{n},p)]^{2}=1.
\label{4.5}
\end{equation}
To evaluate the integral in Eq.\ (\ref{4.5}), we use Eq.\
(\ref{3.38}). This yields
\begin{eqnarray}
\int_{0}^{\infty}\mathrm{d}p\:p^{N-1}
[F_{n-l-1,l}(E_{n},p)]^{2} &=& 2^{2l+N-3}
\left[\Gamma\left(l+\frac{N-1}{2}\right)\right]^{2}
\frac{(n-l-1)!\left(n+\frac{N-3}{2}\right)}{\pi(n+l+N-3)!}\,
q_{n}^{-2}
\nonumber \\
&& \times\int_{-1}^{1}\mathrm{d}\xi_{n}\:
(1-\xi_{n}^{2})^{l+N/2-1}(1+\xi_{n})
\left[C_{n-l-1}^{(l+(N-1)/2)}(\xi_{n})\right]^{2},
\nonumber \\
&&
\label{4.6}
\end{eqnarray}
with
\begin{equation}
q_{n}=\frac{Zq_{\mathrm{B}}}{n+\frac{N-3}{2}}.
\label{4.7}
\end{equation}
It is evident that in the integrand on the right-hand side of Eq.\
(\ref{4.6}) one may replace the factor $1+\xi_{n}$ by $1$ without
altering the value of the integral. Applying then the formula
(\ref{3.33}) and imposing a further restriction on $A$ to be
positive, we obtain
\begin{equation}
A=\sqrt{2}\,q_{n}.
\label{4.8}
\end{equation}
Hence, we conclude that the bound-state momentum-space eigenfunctions
of the Schr{\"o}dinger--Coulomb problem in $\mathbb{R}^{N}$,
$N\geqslant2$, associated with the energy levels (\ref{4.1}) and
orthonormal in the sense of
\begin{equation}
\int_{\mathbb{R}^{N}}\mathrm{d}^{N}\boldsymbol{p}\:
\Phi_{nl\eta}^{*}(\boldsymbol{p})\Phi_{n'l'\eta'}(\boldsymbol{p})
=\delta_{nn'}\delta_{ll'}\delta_{\eta\eta'},
\label{4.9}
\end{equation}
may be chosen in the form
\begin{eqnarray}
\Phi_{nl\eta}(\boldsymbol{p}) 
&=& 2^{N-1/2}\Gamma\left(l+\frac{N-1}{2}\right)
\sqrt{\frac{(n-l-1)!\left(n+\frac{N-3}{2}\right)}{\pi(n+l+N-3)!}}
\nonumber \\
&& \times\,q_{n}^{N/2+1}\frac{(4q_{n}p)^{l}}
{(q_{n}^{2}+p^{2})^{l+(N+1)/2}}\,C_{n-l-1}^{(l+(N-1)/2)}
\left(\frac{q_{n}^{2}-p^{2}}{q_{n}^{2}+p^{2}}\right)
Y_{l\eta}^{(N-1)}(\boldsymbol{n}_{p}).
\label{4.10}
\end{eqnarray}
The degeneracy of the energy level $E_{n}$ is, obviously, the same as
that of the Sturmian eigenvalue $\lambda_{n}$, and is given by Eq.\
(\ref{3.45}).

With Eqs.\ (\ref{4.1}) and (\ref{4.10}) being obtained, our task to
determine bound-state eigensolutions to the Schr{\"o}dinger--Coulomb
problem in the $N$-dimensional Euclidean momentum space is
accomplished.
\section*{Acknowledgments}
I thank Dr.\ Howard S.\ Cohl for drawing my attention to his work
\cite{Cohl11}.
%
%
\appendix
\section*{Appendix: The Fourier transform of
$V(\boldsymbol{r})=Ar^{-\nu}$, $0<\nu<N$, in $\mathbb{R}^{N}$}
\label{A}
\setcounter{equation}{0}
\setcounter{section}{1}
In this Appendix, we shall follow closely Ref.\ \cite[Sec.\
5.9]{Lieb01} (for an alternative approach, cf.\ Ref.\
\cite[Appendix 3]{Levy50}). We wish to evaluate the Fourier transform
\begin{equation}
U(\boldsymbol{p})=\frac{1}{(2\pi\hbar)^{N/2}}
\int_{\mathbb{R}^{N}}\mathrm{d}^{N}\boldsymbol{r}\:
\mathrm{e}^{-\mathrm{i}\boldsymbol{p}\cdot\boldsymbol{r}/\hbar}
V(\boldsymbol{r}),
\label{A.1}
\end{equation}
where 
\begin{equation}
V(\boldsymbol{r})=\frac{A}{r^{\nu}}
\qquad (0<\nu<N).
\label{A.2}
\end{equation}
To accomplish the task, at first we observe that
\begin{equation}
\int_{0}^{\infty}\mathrm{d}\xi\:\xi^{\nu/2-1}\mathrm{e}^{-\xi r^{2}}
=r^{-\nu}\int_{0}^{\infty}\mathrm{d}\eta\:\eta^{\nu/2-1}
\mathrm{e}^{-\eta}=\Gamma\left(\frac{\nu}{2}\right)r^{-\nu},
\label{A.3}
\end{equation}
hence, it follows that
\begin{equation}
r^{-\nu}=\frac{1}{\Gamma\left(\frac{\nu}{2}\right)}
\int_{0}^{\infty}\mathrm{d}\xi\:\xi^{\nu/2-1}\mathrm{e}^{-\xi r^{2}}.
\label{A.4}
\end{equation}
If we plug Eq.\ (\ref{A.4}) into Eq.\ (\ref{A.1}), we obtain
\begin{eqnarray}
U(\boldsymbol{p})
&=& \frac{A}{(2\pi\hbar)^{N/2}\Gamma\left(\frac{\nu}{2}\right)}
\int_{\mathbb{R}^{N}}\mathrm{d}^{N}\boldsymbol{r}\:
\mathrm{e}^{-\mathrm{i}\boldsymbol{p}\cdot\boldsymbol{r}/\hbar}
\int_{0}^{\infty}\mathrm{d}\xi\:\xi^{\nu/2-1}\mathrm{e}^{-\xi r^{2}}
\nonumber \\
&=& \frac{A}{(2\pi\hbar)^{N/2}\Gamma\left(\frac{\nu}{2}\right)}
\int_{0}^{\infty}\mathrm{d}\xi\:\xi^{\nu/2-1}
\int_{\mathbb{R}^{N}}\mathrm{d}^{N}\boldsymbol{r}\:
\mathrm{e}^{-\mathrm{i}\boldsymbol{p}\cdot\boldsymbol{r}/\hbar}
\mathrm{e}^{-\xi r^{2}}.
\label{A.5}
\end{eqnarray}
The integral over $\boldsymbol{r}$ is readily found to be
\begin{equation}
\int_{\mathbb{R}^{N}}\mathrm{d}^{N}\boldsymbol{r}\:
\mathrm{e}^{-\mathrm{i}\boldsymbol{p}\cdot\boldsymbol{r}/\hbar}
\mathrm{e}^{-\xi r^{2}}=\left(\frac{\pi}{\xi}\right)^{N/2}
\mathrm{e}^{-p^{2}/4\hbar^{2}\xi},
\label{A.6}
\end{equation}
so that Eq.\ (\ref{A.5}) becomes
\begin{equation}
U(\boldsymbol{p})=\frac{A}{(2\hbar)^{N/2}\Gamma(\frac{\nu}{2})}
\int_{0}^{\infty}\mathrm{d}\xi\:\xi^{(\nu-N)/2-1}
\mathrm{e}^{-p^{2}/4\hbar^{2}\xi},
\label{A.7}
\end{equation}
and further, after switching to the integration variable
$\eta=p^{2}/4\hbar^{2}\xi$,
\begin{equation}
U(\boldsymbol{p})=A\frac{(2\hbar)^{N/2-\nu}}
{\Gamma\left(\frac{\nu}{2}\right)}\frac{1}{p^{N-\nu}}
\int_{0}^{\infty}\mathrm{d}\eta\:\eta^{(N-\nu)/2-1}\mathrm{e}^{-\eta}.
\label{A.8}
\end{equation}
The remaining integral is elementary and we eventually arrive at
\begin{equation}
U(\boldsymbol{p})
=A\frac{(2\hbar)^{N/2-\nu}\Gamma\left(\frac{N-\nu}{2}\right)}
{\Gamma\left(\frac{\nu}{2}\right)}\frac{1}{p^{N-\nu}}.
\label{A.9}
\end{equation}
In the case of the Coulomb potential
\begin{equation}
V(\boldsymbol{r})=-\frac{Ze^{2}}{(4\pi\epsilon_{0})r},
\label{A.10}
\end{equation}
from Eq.\ (\ref{A.9}) we find that
\begin{equation}
U(\boldsymbol{p})=-\frac{Ze^{2}}{(4\pi\epsilon_{0})}
\frac{(2\hbar)^{N/2-1}\Gamma\left(\frac{N-1}{2}\right)}{\sqrt{\pi}}
\frac{1}{p^{N-1}}.
\label{A.11}
\end{equation}
%
%

%
\end{document}